\documentclass[12pt]{article}
\usepackage{graphicx}
\usepackage{multirow}

\usepackage[T1]{fontenc}
\usepackage[sc]{mathpazo}
\usepackage{amsmath}
\usepackage{amssymb}
\usepackage{enumerate}
\usepackage{amsthm}
\usepackage{amsfonts,mathrsfs}

\usepackage{hyperref}
\hypersetup{pdfpagemode=UseNone}

\newcommand{\tinyspace}{\mspace{1mu}}

\newcommand{\abs}[1]{\left\lvert\tinyspace #1 \tinyspace\right\rvert}

\newenvironment{mylist}[1]{\begin{list}{}{
    \setlength{\leftmargin}{#1}
    \setlength{\rightmargin}{0mm}
    \setlength{\labelsep}{2mm}
    \setlength{\labelwidth}{8mm}
    \setlength{\itemsep}{0mm}}}
    {\end{list}}

\newcommand{\Pa}[1]{\left(#1\right)}

\newcommand{\fontmapset}{\mathbf} 

\newcommand{\Mset}[2]{\fontmapset{#1}\Pa{#2}}
\newcommand{\Lin}[1]{\Mset{L}{#1}}

\theoremstyle{definition}

\numberwithin{equation}{section}

\newcounter{questionnumber}

\begin{document}

\title{\Large\bf A Rogue-Langmuir-type traveling wave continuous solution
to nonlinearly dispersive Schr\"{o}dinger equation using a
polynomial expansion scheme}

\author{Karem Boubaker\footnote{E-mail:mmbb11112000@yahoo.fr}\\[1mm]
{\it\small Unit\'{e} de Physique des dispositifs \`{a}
semi-conducteurs, Tunis El Manar University, 2092 Tunis, Tunisia}
\\Lin Zhang\footnote{E-mail: godyalin@163.com}\\[1mm]
{\it\small Institute of Mathematics, Hangzhou Dianzi University,
Hangzhou 310018, P.R.~China}}

\date{}
\maketitle \mbox{}\hrule\mbox\\
\begin{abstract}

In this paper, traveling wave solutions to the nonlinearly
dispersive Schr\"{o}dinger equation are given in the case of
one-dimensional non-relativistic electron confined to a cylindrical
quantum well. Investigations gave evidence to the possibility of
implementing continuous solutions for a quantum-based problem.
\end{abstract}
\mbox{}\hrule\mbox\\~\\

\section{Introduction}

The well-known nonlinearly dispersive Schr\"{o}dinger equation
\cite{1Davvdova}-\cite{8Wang}, described as:
\begin{eqnarray}\label{eq:1}
\mathrm{i}\frac{\partial u}{\partial t} +
\frac{\partial^2\Pa{u\abs{u}^{n-1}}}{\partial x^2}+\mu u
\abs{u}^{m-1}=0,
\end{eqnarray}
where $u$ is the unknown function which determines the probability
distribution, $\mu$ is a given parameter and $m$ and $n$ are
positive integers  and denote the intensity of the nonlinear term.
This equation arises from the research of nonlinear wave propagation
in dispersive and inhomogeneous media. It has been also encountered
in problems of plasma physics,   hydrodynamics, self trapping of
light with    formation  of  spatiotemporal solitons, evolution of
slowly varying electromagnetic field  as well as early studies of
pulses in optical wave fibers \cite{5Davvdova}-\cite{12Gagnon}.

Exact and analytical solutions for  nonlinearly dispersive
Schr\"{o}dinger equation have attracted considerable attention
\cite{9Pushkarov}-\cite{16Palacios}. Several attempts yielded
families of exact analytical solutions which were obtained  using
elementary functions \cite{11Tanev}-\cite{14Bimbaum}.

In the present work, a polynomial expansion scheme is performed in
order to obtain Rogue-Langmuir-type traveling wave solution to
Eq.~\eqref{eq:1}. This paper is organized as follows. In
Section~\ref{sect:2}, the resolution protocol is presented along
with the studied system patterns. In Section~\ref{sect:3}, plots of
the solutions are shown ad discussed. Last section is the
conclusion.

\section{Resolution protocol}\label{sect:2}

\begin{figure}[htbp]
\centering
\includegraphics[height=3in,width=5in]{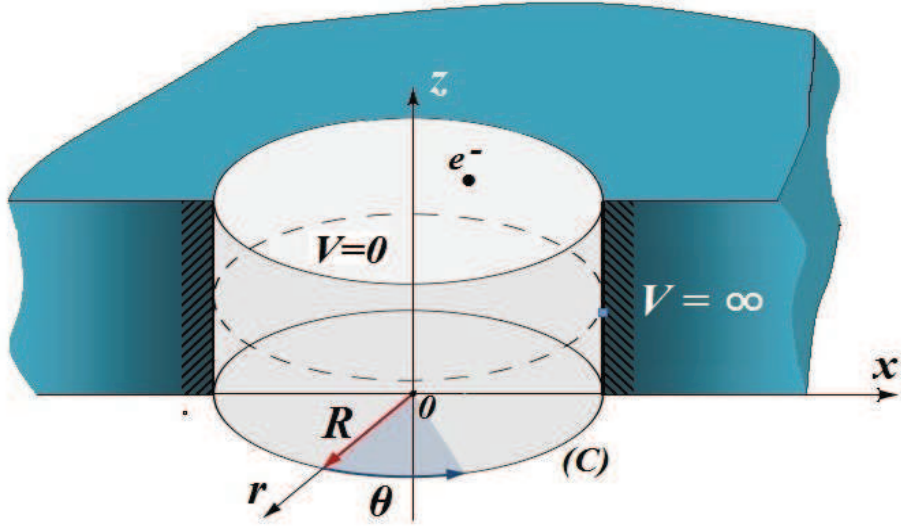}
\caption{Non-relativistic electron moving inside a cylindrical quantum well}\label{fig:1}
\end{figure}

Schr\"{o}dinger's equation is introduced here in the case of a
one-dimensional non-relativistic electron $e$- of mass $m$, moving
inside a cylindrical quantum well $(C)$ of radius $R$
(Fig.~\ref{fig:1}).
The potential in the whole space is defined as
\begin{eqnarray}\label{eq:2}
\begin{cases}
V(r) = 0,& r\in(0,R)\\
V(r) = \infty,& r\in(R,\infty)
\end{cases}
\end{eqnarray}

That is to say, there is an infinitely high cylindrical infinite
R-radius envelop at walls at $r = R$ and the particle is trapped in
the region $r\in[0,R]$.

Schr\"{o}dinger's equation in cylindrical co-ordinate system for the
non-relativistic electron in quantum well is
\begin{eqnarray}\label{eq:3}
\mathrm{i}\hbar\frac{\partial u(r,t)}{\partial t} +
\frac{\hbar^2}{2m}\frac{\partial^2\Pa{u(r,t)\abs{u(r,t)}^{n-1}}}{\partial
x^2} = V(r)u(r,t)\abs{u(r,t)}^{m-1}.
\end{eqnarray}

Since it is the matter of finding a Rogue-Langmuir traveling wave
solution of Eq.~\eqref{eq:1}, we introduce the wave variable
$\theta$, so that
\begin{eqnarray}\label{eq:4}
x = -\frac{\mathrm{i}\hbar}{p}\theta + Et.
\end{eqnarray}
We have consequently:
\begin{eqnarray}\label{eq:5}
\begin{cases}
\frac{\partial}{\partial x}(\cdot) &= -\frac{\mathrm{i}\hbar}{p}\frac{\partial}{\partial \theta}(\cdot)\\
\frac{\partial^2}{\partial^2 x}(\cdot) &= -\frac{\mathrm{i}\hbar}{p}\frac{\partial^2}{\partial^2 \theta}(\cdot)\\
\frac{\partial}{\partial t}(\cdot) &= -\frac{\mathrm{i}\hbar}{pE}\frac{\partial}{\partial \theta}(\cdot)\\
\end{cases}
\end{eqnarray}
so that Eq.~\eqref{eq:1} becomes, for $n = 2$  and $m = 3$:
\begin{eqnarray}\label{eq:6}
\begin{cases}
\Pa{\frac p\hbar}^2 \frac{d^2f(\theta)}{d\theta^2} + \Pa{\frac
E\hbar + \Pa{\frac p\hbar}^2}f(\theta) - V(r)f^3(\theta) = 0&\\
f(\theta) = u(x,t)\exp\Pa{-\frac{\mathrm{i}}{\hbar}(px-Et)} =
u(x,t)e^{-\theta} &~
\end{cases}
\end{eqnarray}
Since $\abs{u}^2$ represents the probability of finding the electron
anywhere, we have the trivial condition $\abs{u}\equiv 0$ for
$r\in(R,\infty)$. According to the BPES principles, the expression
of the unknown term of the traveling wave solution is proposed as
following
\begin{eqnarray}\label{eq:7}
f(\theta) = \frac1{2N_0} \sum^{N_0}_{k=1}\widetilde{\xi_k}\cdot
B_{4k}(\theta\mu_k),
\end{eqnarray}
where $B_{4k}$ are the $4k$-order Boubaker polynomials, $\mu_k$ are
$B_{4k}$ minimal positive roots \cite{17Milgram}-\cite{36Rahmanov},
$N_0$ is a prefixed integer, and $\widetilde{\xi_k}(k=1,\ldots,N_0)$
are unknown pondering real coefficients.

The main advantage of these formulations Eq.~\eqref{eq:7} is the
fact of verifying boundary conditions, at the earliest stage of
resolution protocol thanks to the properties of the Boubaker
polynomials \cite{19Slama}-\cite{33Kumar}:
\begin{eqnarray}\label{eq:8}
\begin{cases}
\sum^N_{q=1}B_{4q}(x)|_{x=0} &= -2N\neq0\\
\sum^N_{q=1}B_{4q}(x)|_{x=\mu_q} &= 0\\
\end{cases}
\end{eqnarray}
and
\begin{eqnarray}\label{eq:9}
\begin{cases}
\sum^N_{q=1}\frac{dB_{4q}(x)}{dx}|_{x=0} &=0\\
\sum^N_{q=1}\frac{dB_{4q}(x)}{dx}|_{x=\mu_q} &= \sum^N_{q=1} H_q\\
\end{cases}
\end{eqnarray}
with
$$
H_n=B_{4n}(\mu_n) =
\frac{4\mu_n(2-\mu^2_n)\sum^n_{q=1}B^2_{4q}(\mu_n)}{B_{4(n+1)}(\mu_n)}
+ 4\mu^3_n.
$$

Thanks to the properties expressed by Eqs.~\eqref{eq:8},
\eqref{eq:9}, boundary conditions are trivially verified in advance
to resolution process. Eq.~\eqref{eq:6} becomes, for the given
potential expression in Eq.~\eqref{eq:2}:
\begin{eqnarray}\label{eq:10}
\Pa{\frac p\hbar}^2 \sum^{N_0}_{k=1}
\widetilde{\xi_k}\mu^2_k\frac{d^2B_{4k}(\theta\mu_k)}{d\theta^2} +
\Pa{\frac E\hbar + \Pa{\frac p\hbar}^2} \sum^{N_0}_{k=1}
\widetilde{\xi_k}B_{4k}(\theta \mu_k)=0.
\end{eqnarray}

The BPES solution is obtained by determining the non-null set of
coefficients $\widetilde{\xi_k}^{\mathrm{(sol.)}}(k=1,\ldots,N_0)$
that minimizes the absolute functional $\Omega_{N_0}$:
\begin{eqnarray}\label{eq:11}
\Omega_{N_0} = \abs{\sum^{N_0}_{k=1}\mu^2_k
\widetilde{\xi_k}^{\mathrm{(sol.)}}\Lambda_k +
\frac1{2N_0}\sum^{N_0}_{k=1}
\widetilde{\xi_k}^{\mathrm{(sol.)}}\Lambda'_k}
\end{eqnarray}
with
$$
\Lambda_k = \frac{p^2}{2\hbar^2
N_0}\oint_{(C)}\frac{d^2B_{4k}(\theta\mu_k)}{d\theta^2}d\theta,\quad
\Lambda'_k = \frac1{2N_0}\Pa{\frac E\hbar + \Pa{\frac
p\hbar}^2}\oint_{(C)}B_{4k}(\theta\mu_k)d\theta.
$$
Finally the solution of Eq.~\eqref{eq:3} is
\begin{eqnarray}\label{eq:12}
u^{\mathrm{(sol.)}}(x,t) = \frac1{2N_0}\sum^{N_0}_{k=1}
\widetilde{\xi_k}^{\mathrm{(sol.)}}B_{4k}\Pa{\frac{\mathrm{i}}{\hbar}(px-Et)\mu_k}\exp\Pa{-\frac{\mathrm{i}}{\hbar}(px-Et)}
\end{eqnarray}
where $x\in[0,R],t\in[0,t_m], t_m = \frac{2\pi\hbar}{E}$.

The main advantage of this solution is the fact that, oppositely to
most on classical solutions \cite{37Fabien}-\cite{40Fujioka},  no
quantification is imposed to the couple $(p,E)$. The solution is
hence uni-modal and piecewise continuous. Moreover, convergence is
obtained for moderate values of $N_0$, since, as mentioned above,
boundary conditions were verified in advance to resolution process.

\section{Solution plots and patterns}\label{sect:3}

\begin{figure}[htbp]
\centering
\includegraphics[height=3in,width=6in]{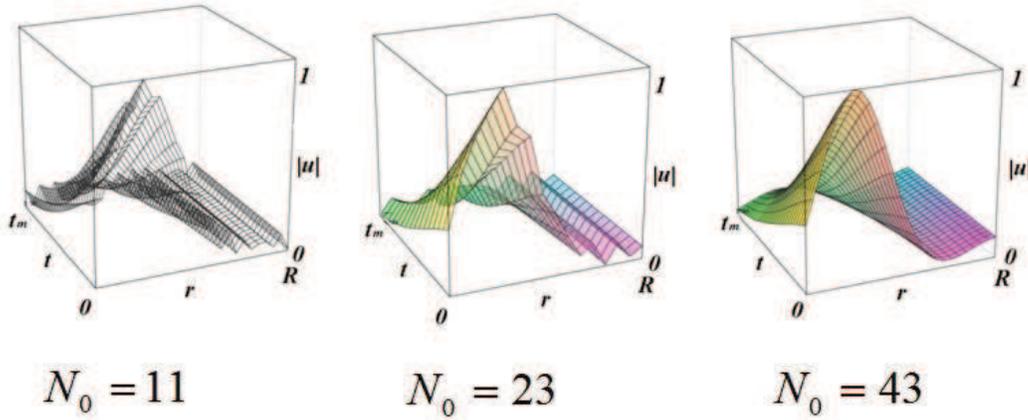}
\caption{Solution convergence patterns(plots for
$N_0=11,23,43$)}\label{fig:2}
\end{figure}

\begin{figure}[htbp]
\centering
\includegraphics[height=3in,width=4in]{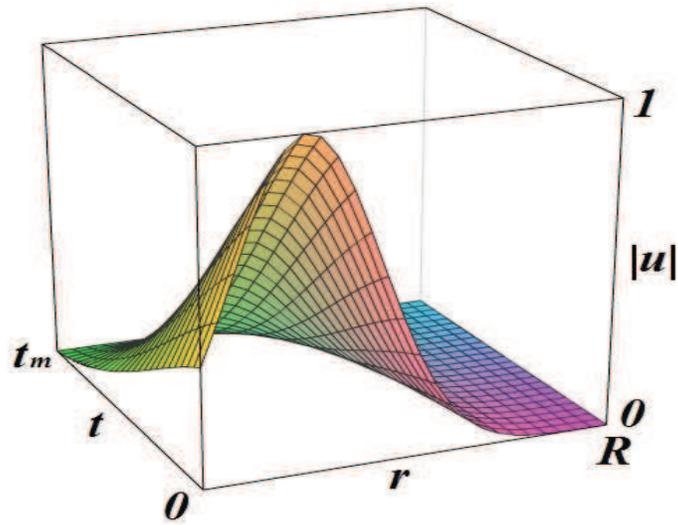}
\caption{Solution plots for $N_0>57$}\label{fig:3}
\end{figure}

Fig.~\ref{fig:2} shows that plots of the solution, for increasing
values of $N_0(N_0=11,23,43)$, while Fig.~\ref{fig:3} corresponds to
the convergent solution modulus, obtained for $N_0>57$. All the
solutions have been represented with $[0,1]$ and $[0,t_m]$ as space
and time ranges, respectively.

\begin{figure}[htbp]
\centering
\includegraphics[height=3in,width=5in]{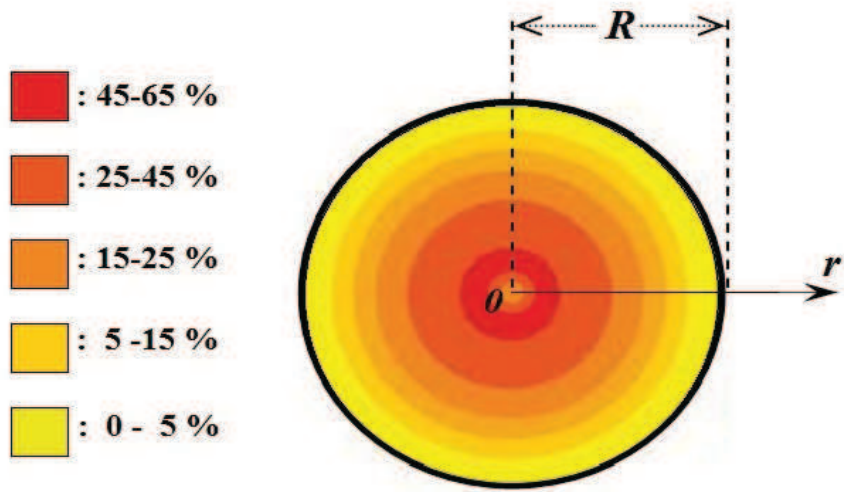}
\caption{Probability distribution $\Pa{\abs{u}^2}$ within the
cylinder $(C)$}\label{fig:4}
\end{figure}

It may be appropriate to point out that Eq.~\eqref{eq:3} is derived
for short amplitude quasi-stationary  slow motion describing the
Rogue-Langmuir pondero-motive force. Most of classical solutions,
which describe a classical-type particle motion under the action of
such forces, consist of linear sums of wave functions corresponding
to different energies \cite{41Amusia}. The present solution accounts
for the trapping of such waves in an infinite well, and oppositely
to many other results, it concentrates the electron energy into a
small region near at the vicinity of the central zone
(Fig.~\ref{fig:4}). This paradox can be explained by the nonlinear
properties of the medium as well as the abrupt potential
discontinuity at the envelope $r=R$.

Fig.~\ref{fig:4} presents the probability distribution within the
cylinder $(C)$. It monitors a typical single energy wave function
having a static probability distribution.

\section{Conclusion}

In summary, we have proposed piecewise continuous and uni-modal
Rogue-Langmuir-type traveling wave solution to the well known
Schr\"{o}dinger equation. The performed polynomial scheme has
ensured the verification of boundary condition in advance to
resolution process. The obtained solutions have been expressed in
terms of wave function modulus and presented the singular advantage
of imposing no quantification for both particle momentum and energy
oppositely to most classical solutions. The convergence of the
protocol has been discussed and enhanced.



\end{document}